\documentclass[12pt]{article}
\usepackage{epsfig}

\begin{document}
\begin{center}
\Large\bf
The detection of Class I methanol masers towards regions
of low-mass star formation.
\large\rm

\bigskip
S. V. Kalenskii, V. G. Promyslov, V. I. Slysh

\normalsize\medskip
{\em Astro Space Center of P.N. Lebedev's Physical Institute}

\large\bigskip
P. Bergman, A. Winnberg

\normalsize\medskip
{\em Onsala Space Observatory}
\end{center}

\begin{abstract}
Six young bipolar outflows in regions of low-to-intermediate-mass
star formation were observed in the $7_0-6_1A^+$, $8_0-7_1A^+$,
and $5_{-1}-4_0E$ methanol lines at 44, 95, and 84~GHz,
respectively. Narrow features were detected
towards NGC~1333IRAS4A, HH~25MMS, and L1157 B1. Flux densities of
the detected lines are no higher than 11 Jy, which is much lower
than the flux densities of strong maser lines in regions of
high-mass star formation. Analysis shows that most likely the
narrow features are masers.
\end{abstract}

\section*{Introduction}
Bright and narrow maser lines of methanol (CH$_3$OH) have been
found towards many star-forming regions. According to the
classification by Menten~\cite{menten91}, all methanol masers can
be divided into two classes, I and II. The Class I masers emit in
the transitions $7_0-6_1A^+$ at 44~GHz, $4_{-1}-3_0E$ at 36~GHz,
$5_{-1}-4_0E$ at 84~GHz etc, the Class II masers emit in the
transitions $5_1-6_0A^+$ at 6.7~GHz, $2_0-3_{-1}E$ at 12~GHz, and
in the series of transitions $J_0-J_{-1}E$ at 157~GHz etc.
Class II masers are often associated with UC HII regions or IR sources.
According to current views, Class I masers are pumped by
collisions, and Class II masers are pumped by external radiation.

As a result of a number of surveys in the Class I lines
$7_0-6_1A^+$, $4_{-1}-3_0E$, $5_{-1}-4_0E$ etc., more than one
hundred Class I masers have been found in high-mass star-forming
regions. At the same time, Bachiller~et~al.~\cite{bach90} and
Kalenskii~et~al.~\cite{kalen92} observed a large number of both
high-mass and low-mass young stars in the $7_0-6_1A^+$ line and
did not find any maser towards low-mass stars. Therefore it is
generally considered that the Class I masers are related to
massive stars. An exception is the detection by
Kalenskii~et~al.~\cite{kalen01} of a narrow, probably, maser
feature in the Class I lines $5_{-1}-4_0E$ and $8_0-7_1A^+$
towards the blue wing of the bipolar outflow driven by a young
low-mass star L1157-mm. Gas in this direction is shock heated to a
temperature of about 100~K, and the abundances of methanol,
ammonia, formaldehyde, and many other molecules are enhanced due to
grain mantle evaporation.

\begin{table}
\caption{The main parameters of the observed lines and those of
the telescope at the line frequencies.}
\label{freqtabl}
\bigskip
\begin{center}
\begin{tabular}{|l|c|c|c|c|}
\noalign{\medskip}
\hline\noalign{\smallskip}
Transition       &Frequency   &Line   & HPBW$^2$ & Jy/K\\
                 &  (GHz)    &strength$^1$&$('')$    &    \\
\noalign{\smallskip}
\hline\noalign{\smallskip}

$7_0-6_1A^+$     & 44.069476  & 2.8609& 82   &20.5\\
$5_{-1}-4_0E$    & 84.521206  & 1.4115& 44   &22\\
$8_0-7_1A^+$     & 95.169516  & 3.3377& 39   &25\\
$2_{-1}-1_{-1}E$ & 96.739393  & 1.5   & 39   &25\\
$2_0-1_0A+$      & 96.741377  & 2     & 39   &25\\
$2_0-1_0E$       & 96.744549  & 2     & 39   &25\\
$2_1-1_1E$       & 96.755507  & 1.5   & 39   &25\\
\hline
\end{tabular}
\end{center}

\medskip
$^1$--from Lees et al.~\cite{lees}

$^2$--half-power beamwidth
\end{table}

The reason for nondetection of masers at 44~GHz by
Bachiller~et~al.~\cite{bach90} and Kalenskii~et~al.~\cite{kalen92}
might be an insufficiently high sensitivity (about 10~Jy) or poor
choice of either star-forming regions and/or observing positions.
In particular, it is reasonable to assume that methanol masers are
formed towards the wings of bipolar outflows, where the abundance
of methanol is enhanced~\cite{plammen}. However, in the survey by
Bachiller~et~al.~\cite{bach90} the telescope was often pointed
towards $IRAS$ objects, and in the survey by
Kalenskii~et~al.~\cite{kalen92} it was pointed towards these
objects almost without exception. $IRAS$ objects spatially
coincide with central sources of outflows rather than with their
wings. The low-mass objects that were observed by Bachiller~et~al.
and Kalenskii~et~al. are located relatively close to the Sun,
usually at a distance of 500~pc or closer. The regions, where the
gas is shock heated, are often located approximately $1'$ away
from the central sources or even further. The masers, related to
these regions, would fall outside the $2'$ wide main beam of the
telescope, used in the observations by Bachiller~et~al. and
Kalenskii~et~al. The regions of massive star formation are much
more rare objects than the regions of low mass star formation.
Most of them are located at several kpc from
the Sun, and their angular sizes are usually less than $1'$.
Therefore, in the single-dish observations the masers fall within
the antenna beams and tend to be detected independently of their
exact locations in star-forming regions. For example, the masers
in a relatively nearby (2 kpc from the Sun) region of massive star
formation L379 are located towards a bipolar outflow wing;
however, they fell into the main beam of the telescope when
Kalenskii~et~al.~\cite{kalen92} observed the central source,
$IRAS\;18265-1517$. The closest maser is $30''$ offset from this
object, which corresponds to a linear distance of
0.3~pc\footnote{For simplicity, we call "distance" the projection
of distance on the plane of the sky.}. If L379 were located at 500
pc from the Sun, all angular distances would increase by a factor
of 4. The masers would be $120''$ offset from the $IRAS$ object
and would not fall into the telescope beam.

\begin{table}
\caption{Gaussian parameters of the observed single lines. The
notation of the lines: I,~$7_0-6_1A^+$; II,~$8_0-7_1A^+$;
III,~$5_{-1}-4_0E$.} \label{gauss}
\begin{tabular}{|l|r|c|r|r|r|r|}
\noalign{\medskip}
\hline\noalign{\smallskip}
Source    &R.A.$_{B1950}$&Line &$\int T^*_A dV\;\;$&$V_{\rm LSR}\;\;$ & FWHM       & $T^*_A$\\
          &DEC.$_{B1950}$ &   &(K$\cdot$ km s$^{-1}$)
                                          &(km s$^{-1}$) &(km s$^{-1}$)& (K)\\
\noalign{\smallskip}
\hline\noalign{\smallskip}
NGC~1333IRAS2 & 03 25 54.5&I&1.33(0.06)$^1$&10.83(0.17) & 7.88(0.45) & 0.16 \\
             & 31 04 00   &II & 0.98(0.08)& 9.97(0.28) & 8.38(0.85) & 0.11 \\
         &            &III& 5.32(0.09)&10.14(0.04) & 5.46(0.10) & 0.92 \\
NGC~1333IRAS4A & 03 26 04.8 & I &0.16(0.02) & 6.65(0.13) & 1.59(0.22) & 0.095 \\
             & 31 03 13   & I &0.09(0.02) & 7.51(0.02) & 0.33(0.05) & 0.25 \\
             &            & II& 1.01(0.11)& 4.38(0.78) & 8.93(1.49) & 0.11 \\
         &            & II& 0.44(0.06)& 7.32(0.03) & 0.94(0.14) & 0.45 \\
             &            &III& 1.60(0.12)& 3.78(0.16) & 7.19(0.31) & 0.21 \\
             &            &III& 0.16(0.04)& 6.95(0.04) & 0.67(0.13) & 0.23 \\
             &            &III& 0.64(0.03)& 6.42(0.10) & 2.31(0.17) & 0.26 \\
NGC~1333IRAS4B & 03 26 06.5 & I &0.32(0.03) & 7.10(0.07) & 1.83(0.21) & 0.16 \\
             & 31 02 51   & II& 1.15(0.07)& 5.36(0.30) & 9.31(0.57) & 0.12 \\
             &            & II&0.16(0.04)$^2$&6.08(0.07)&0.76(0.20) & 0.20 \\
             &            & II& 0.33(0.03)& 7.16(0.04) & 0.76(0.08) & 0.40 \\
             &            &III& 2.70(0.09)& 4.72(0.15) & 8.77(0.29) & 0.29 \\
             &            &III& 0.73(0.05)& 6.86(0.04) & 1.59(0.10) & 0.43 \\
HH~25MMS      & 05 43 34.0 & I &0.27(0.02) &10.42(0.01) & 0.48(0.04) & 0.52 \\
             &$-$00 15 20 & II&0.81(0.11) & 9.97(0.36) & 5.60(0.80) & 0.14 \\
             &            & II&0.31(0.04) &10.14(0.03) & 0.53(0.06) & 0.54 \\
             &            &III&1.63(0.08) &10.53(0.07) & 2.87(0.17) & 0.53 \\
L1157 B1     & 20 38 41.0 & I &0.60(0.03) & 0.69(0.08) & 3.82(0.24) & 0.17 \\
             & 67 50 33   & I &0.12(0.01) & 0.75(0.01) & 0.37(0.03) & 0.31 \\
             &            & II&1.61(0.09)$^1$&$-$0.18(0.15)&5.19(0.33)&0.29 \\
             &            &III&4.61(0.07)$^1$&0.10(0.04)&4.81(0.09) & 0.90 \\
L1157 B2     & 20 38 42.2 & I &0.61(0.09) & 0.04(0.38) &5.29(0.43) & 0.11 \\
             & 67 50 03   & I &0.28(0.11)$^2$&1.53(0.14)& 2.17(0.49)& 0.12 \\
             &            & I &0.04(0.01)$^2$&0.72(0.03)& 0.36(0.08)& 0.11 \\
\noalign{\smallskip}
\hline\noalign{\smallskip}
\end{tabular}

\medskip
$^1$--the line has a non gaussian profile, but the division into 2
or more components is unreliable.

$^2$--marginal detection
\end{table}

There are other regions where Class I masers are located so far
from $IRAS$ objects and bipolar outflow centers that the angular
distances between the masers and these objects would be larger
than $60''$, if the distances to these sources were no more than
200--500~pc. For example, the maser spots in W 33Met are 0.7~pc
away from the nearest known IR object, which would correspond to
$287''$ offset, if the distance to the source were 500~pc. However,
there are regions where the offsets of methanol masers from $IRAS$
objects or bipolar outflow centers are small. For instance, the
maser in M8E is located only 0.02~pc away from the center of a
bipolar outflow~\cite{vm8e}, which, even at a distance as small as
200~pc, would correspond to an offset of only $23''$. The Class I
maser GGD~27 within the errors coincides spatially with the center
of a bipolar outflow~\cite{sl99}. Therefore it is unlikely that
Bachiller~et~al.~\cite{bach90} and Kalenskii~et~al.~\cite{kalen92}
failed to detect any maser towards low-mass stars only because
they observed mostly $IRAS$ objects; probably, the sensitivity of
these surveys also was insufficient.

\begin{table}
\caption{Gaussian parameters of the lines from the series
$2_K-1_K$. The lines were approximated under the assumption that
the widths and radial velocities of all lines are the same. The
second column presents the antenna temperature of the
$2_{-1}-1_{-1}E$ line, the third---that of the $2_0-1_0A^+$ line,
the fourth---that of the $2_0-1_0E$ line, the fifth---that of the
$2_1-1_1E$ line.} \label{gaussth}
\begin{tabular}{|l|c|c|r|r|r|r|}
\noalign{\medskip}
\hline\noalign{\smallskip}
Source       &$T^*_A$&$T^*_A$&$T^*_A$&$T^*_A$& $V_{\rm LSR}\;\;$ & FWHM \\
             & (K)   & (K)   & (K)   & (K)   &(km s$^{-1}$)  &(km s$^{-1}$)\\
\noalign{\smallskip}
\hline\noalign{\smallskip}
NGC~1333IRAS2 &0.45(0.03)&0.42(0.02)&0.15(0.02)&$<$0.06&9.51(0.14)&5.56(0.16)\\
             &0.30(0.03)&0.41(0.03)&$<$0.09   &       &7.53(0.05)&1.42(0.06)\\
NGC~1333IRS4A &0.25(0.05)&0.37(0.05)&0.09(0.04)&$<$0.10&6.22(0.29)&5.31(0.37)\\
             &0.79(0.07)&1.04(0.07)&$<$0.20   &       &7.69(0.03)&1.14(0.04)\\
NGC~1333IRS4B &0.58(0.07)&0.90(0.09)&$<$0.21   &       &7.74(0.07)&1.24(0.07)\\
HH~25MMS      &0.51(0.02)&0.76(0.03)&0.13(0.02)&$<$0.06&10.25(0.05)&3.56(0.07)\\
L1157 B1     &0.75(0.02)&1.10(0.02)&0.28(0.02)&$<$0.06&0.43(0.05) &5.00(0.05)\\
L1157 B2     &0.55(0.02)&0.75(0.02)&0.17(0.02)&$<$0.06&1.30(0.06) &4.00(0.07)\\
\noalign{\smallskip}
\hline\noalign{\smallskip}
\end{tabular}
\end{table}

The detection of methanol masers in relatively nearby sources
could be very important for the maser exploration. Therefore we
observed several nearby bipolar outflows from low-mass stars in
the $7_0-6_1A^+$ line at 7~mm, as well as in the 3mm wave
range\footnote{The observations at 3mm are part of a more extended
survey; the results of this survey will be published elsewhere.},
in the $5_{-1}-4_0E$, $8_0-7_1A^+$ line, and in the series of
$2_K-1_K$ lines. The lines of this series are purely thermal and
never exhibit maser emission. Several objects with enhanced
abundance of methanol were selected from the literature as target
sources.

\section*{Observations}
The observations were performed with the 20-m radio telescope at
Onsala (Sweden). The line frequencies and strengths and the main
parameters of the telescope are presented in Table~\ref{freqtabl}.
A dual beam switching mode with a frequency of 2~Hz and a beam throw
of $11'$ was applied. Pointing errors were checked using
observations of SiO masers and found to be within $5''$. The data
were calibrated using a chopper-wheel method. Data were reduced
with the CLASS package. The line rest frequencies were taken from
Lovas'
database\footnote{http://physics.nist.gov/cgi-bin/micro/table5/start.pl}.

Observations in the $7_0-6_1A^+$ line were carried out on December
6--9, 2004. The main beam efficiency was about 0.5 for elevations
lower than $30^\circ$ and about 0.56 for elevations higher than
$30^\circ$. The half-power beamwidth was $82''$. The system noise
temperature corrected for atmospheric absorption, rearward
spillower, and radome losses varied between 180 and 400~K. An
autocorrelator with a 12.5~kHz (0.085~km~s$^{-1}$ at 44~GHz)
resolution was used as spectrometer.

Observations in the 3mm wave range were carried out in May 2001.
The main beam efficiency and the half-power beamwidth at 84.5~GHz
were 0.6 and $44''$, respectively. A cryogenically cooled
low-noise SIS mixer was used. The system noise temperature
corrected for atmospheric absorption, rearward spillower, and
radome losses varied between 300 and 1000~K. The backend
consisted of an autocorrelator with a 50~kHz resolution (0.177~km
s$^{-1}$ at 84.5~GHz); in parallel, a 256-channel filter
spectrometer with a 250~kHz resolution (0.887~km~s$^{-1}$) was
connected.

\section*{Results}

\begin{figure}
\includegraphics{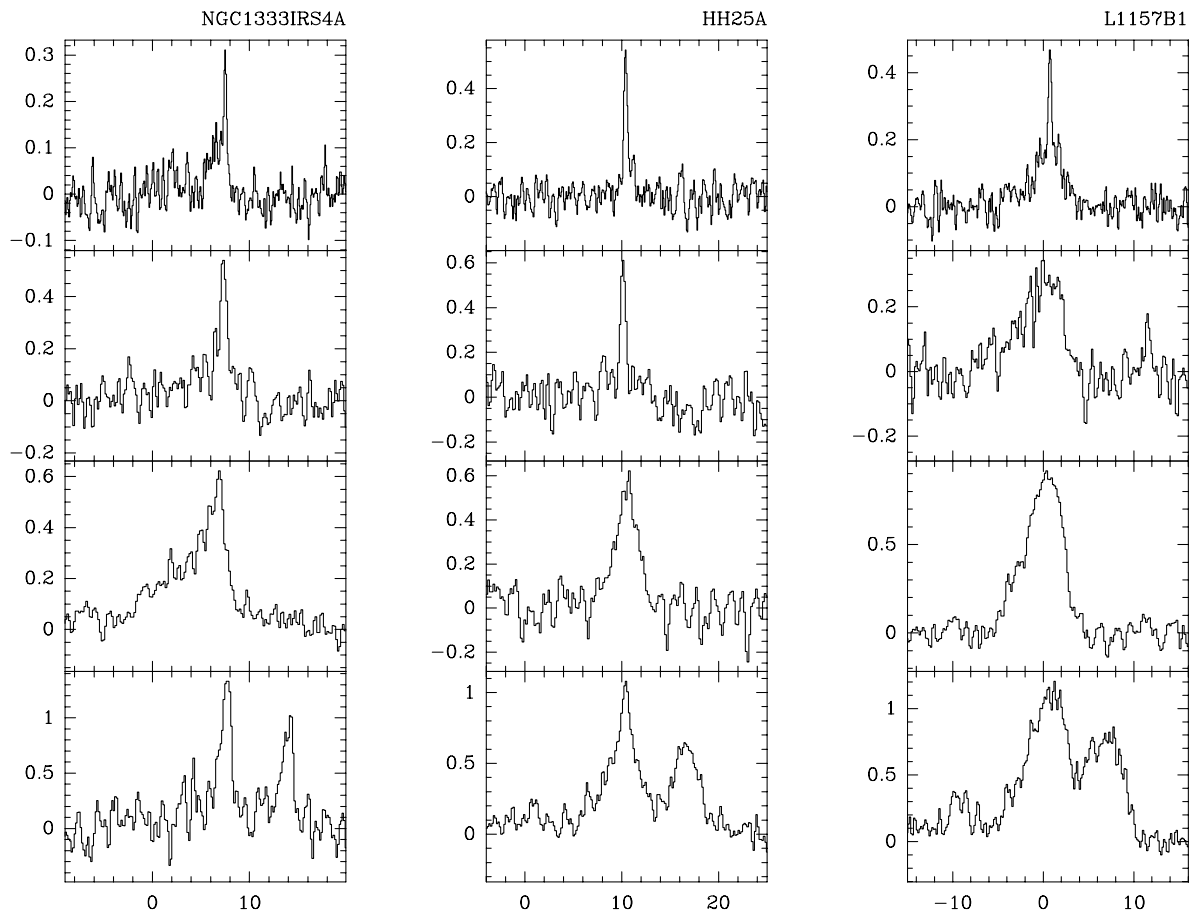}
\caption{Spectra of the sources in which the masers in the
$7_0-6_1A^+$ line were found. The order of the lines shown is as
follows: $7_0-6_1A^+$ (uppermost panel),  $8_0-7_1A^+$,
$5_{-1}-4_0E$, $2_K-1_K$ (lowermost panel). The horizontal axis
gives the radial velocity in km~s$^{-1}$ and the vertical axis
gives the antenna temperature in Kelvins.} \label{mas}
\end{figure}

\begin{figure}
\includegraphics{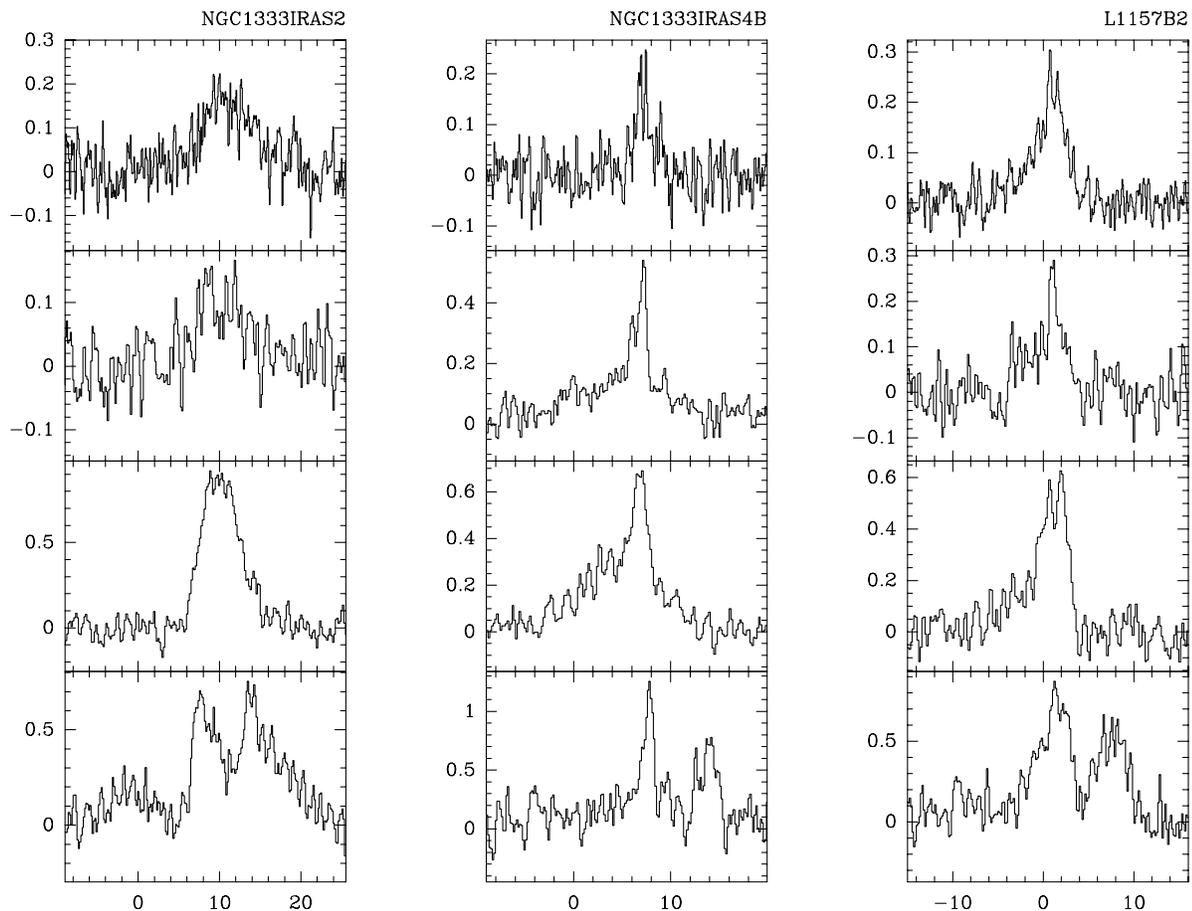}
\caption{Spectra of the sources in which only thermal emission was
detected. The spectra of L1157~B2 at 84 and 95~GHz are taken from
the paper by Kalenskii~et~al.~\cite{kalen01}. The order of the
lines and the X and Y axes are the same as in Fig.~1.}
\label{therm}
\end{figure}

The results are presented in Figures~\ref{mas} and~\ref{therm} and
in Tables~\ref{gauss} and~\ref{gaussth}. Emission at 7~mm was
detected in all six observed regions. Narrow features were found
towards NGC~1333IRAS4A, HH~25MMS, and L1157~B1. Antenna
temperatures of the detected lines do not exceed 0.52~K, which
corresponds to a flux density of 10.7~Jy. Thus, the detected lines
are much weaker than strong maser lines in massive star-forming
regions, which have flux densities up to several hundred Janskys.
In addition, the flux densities of the detected lines are
approximately equal to or lower than the upper limits on flux
density in the surveys by Bachiller~et~al.~\cite{bach90} and
Kalenskii~et~al.~\cite{kalen92}. Thus, the fact that
Bachiller~et~al. and Kalenskii~et~al. did not find any methanol
maser in the $7_0-6_1A^+$ line may be either fully or partly
explained by insufficient sensitivity of these surveys.

\begin{figure}
\vspace{4cm}
\hspace{10mm}
\includegraphics{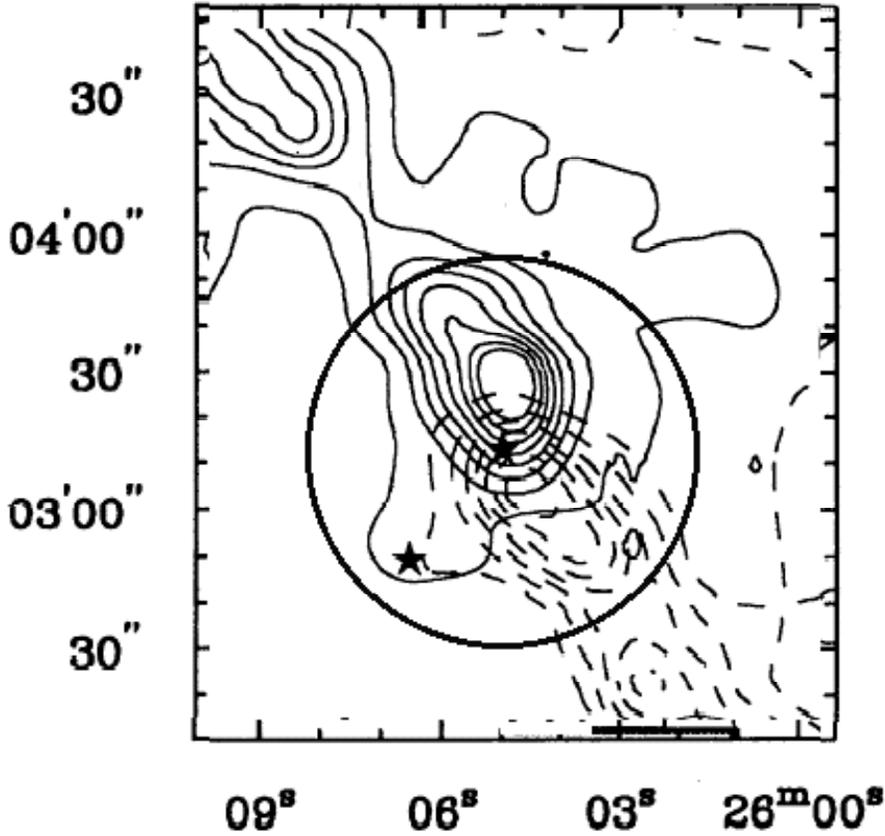}
\vspace{-6cm} \caption{Map of the environment of the IR source
IRAS4A in the cloud NGC 1333. Solid contours: the red wing of the
bipolar outflow; dashed contours: the blue wing of the bipolar
outflow (from the paper by Blake~et~al.~\cite{blake_ngc133}). The
circle of $82''$ in diameter shows the size and position of the
main beam of the telescope during the observations at 7~mm. The
stars denote the sources IRAS-4A (in the center of the circle) and
IRAS-4B (south-west of IRAS-4A).} \label{n1333}
\end{figure}

We observed several regions in the molecular cloud NGC~1333, where
low-to-intermediate-mass stars actively form. A narrow feature at
44~GHz was detected towards the infrared source IRAS4A.
Figure~\ref{n1333} shows that both the central source and peaks of
CO emission in the red and blue wings of the outflow fall into the
antenna beam. Therefore without additional observations with high
spatial resolution we cannot determine to which of these objects
the narrow features are related. Blake~et~al.~\cite{blake_ngc133}
found that the abundances of CS, SiO, and CH$_3$OH in this
direction are enhanced due to grain mantle evaporation. In the
same direction at 84 and 95~GHz we detected features, which are
broader than the narrow feature at 44~GHz, but narrower than the
thermal lines at 96~GHz (Fig.~\ref{mas}); probably, these features
are masers with the absolute values of optical thickness and
hence, line narrowing, less than those at 44~GHz.

\begin{figure}
\vspace{4cm}
\hspace{2cm}
\includegraphics{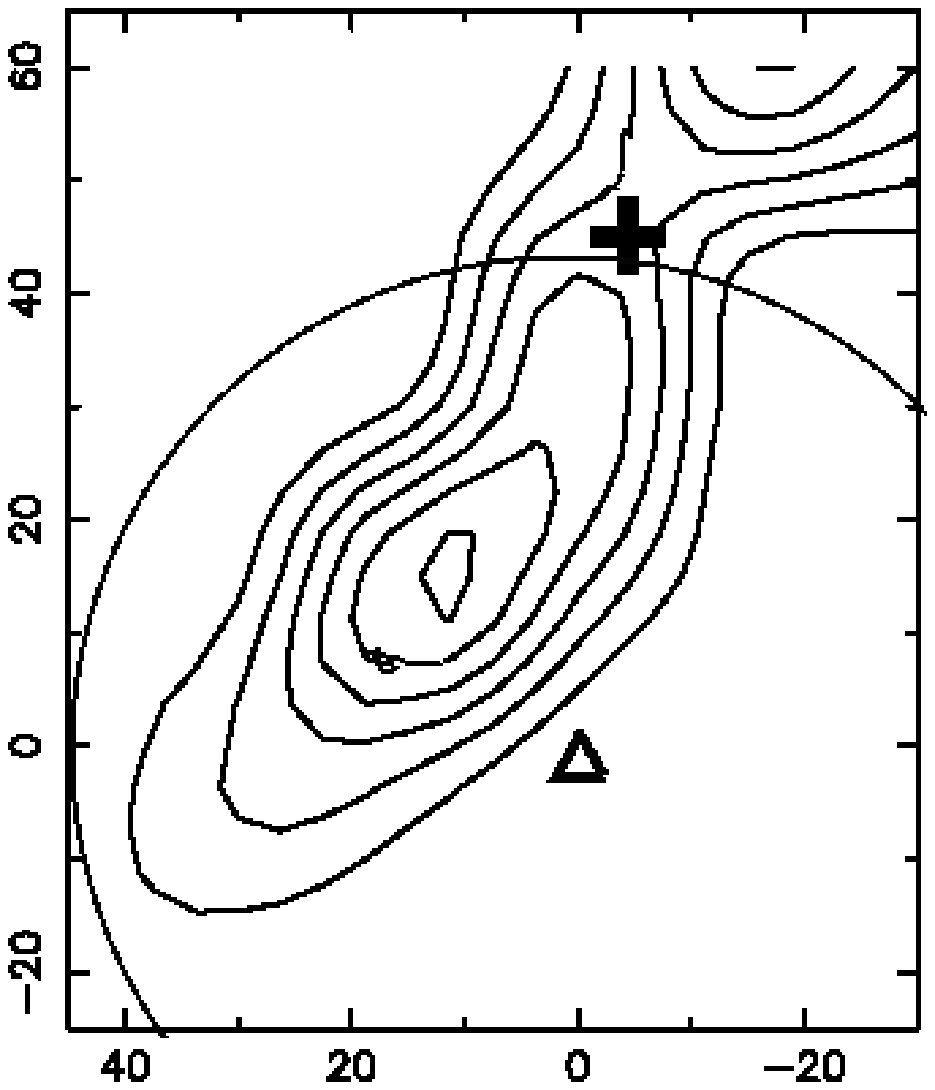}
\vspace{-6cm} \caption{Map of the environment of the submm source 
HH~25MMS. Solid contours show the red wing of the bipolar outflow
in the $J=3-2$ CO line (from the paper by Gibb~and~Davis~\cite{gibbmeth});
the central star is denoted by the cross. The semicircle with the triangle
in the center shows the size and position of the main beam of the
antenna during the observations at 7~mm. The X and Y axes give the
difference in the right ascension and declination (in arcsec) from
the coordinates presented in Table~2.} \label{hh25}
\end{figure}

Narrow lines at 44 and 95~GHz were detected towards the red wing
of a bipolar outflow in the dark cloud L1630 with the central
source HH~25MMS, located south of the IR source $IRAS\;05435-0014$
(Fig.~\ref{hh25}). A broad line was found at 84~GHz. Its width is
close to those of the thermal lines $2_K-1_K$. A high abundance of
methanol in this direction was detected by Gibb and
Davis~\cite{gibbmeth}.

\begin{figure}
\vspace{8cm}
\hspace{-19mm}
\includegraphics{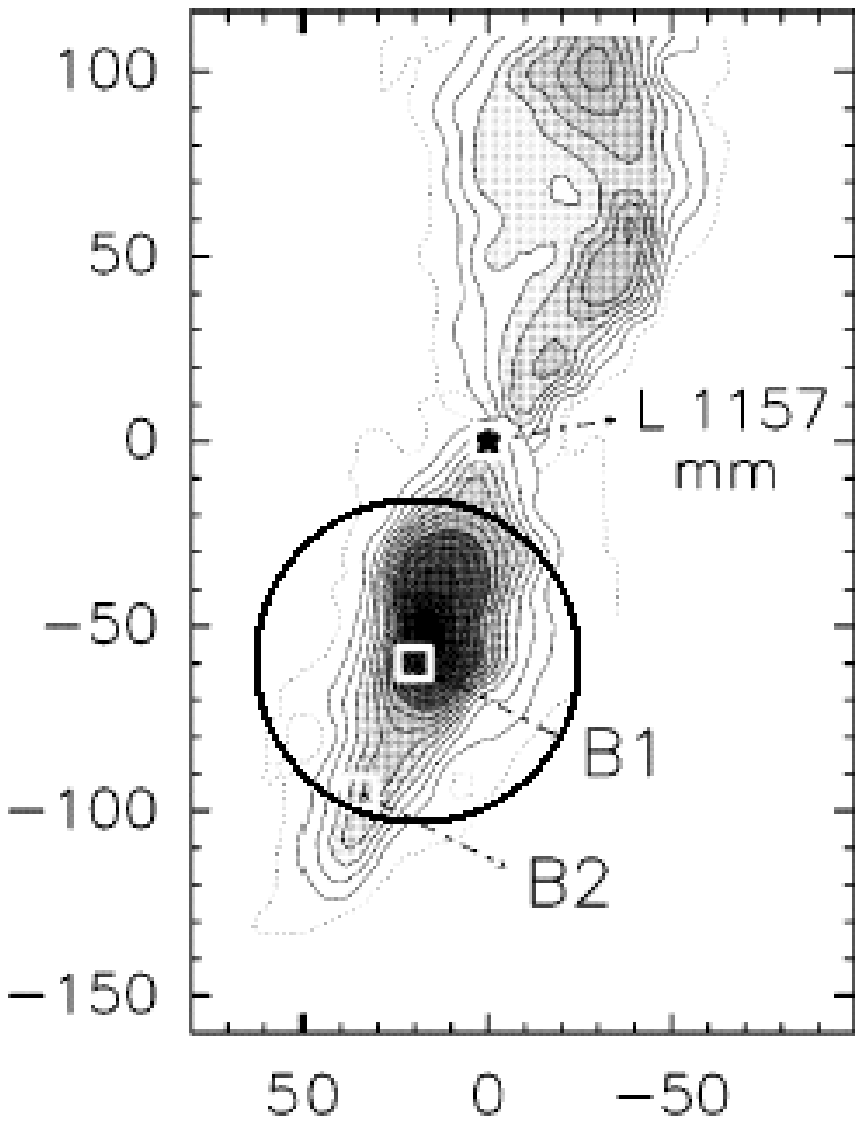}
\vspace{-8cm} \caption{The vicinities of the submm source
L1157-mm. The map of the bipolar outflow is taken from the paper
by Bachiller and P\'erez Guti\'errez~\cite{bach97}. The circle
with the center at B1 demonstrates the size and position of the
main beam of the antenna during the observations at 7~mm.
The X and Y axes give the right ascension and declination offsets (in
arcsec) from the coordinates ($J2000$) $\alpha=20^h 39^m 06^s.19$,
$\delta=68^\circ 02' 15''.9$.} \label{l1157}
\end{figure}

A narrow feature at 44~GHz is also found towards the region B1 in
the blue wing of the bipolar outflow, driven by a class 0 object
(according to the Andre~et~al~\cite{andre} classification) L1157-mm
in the molecular cloud L1157 (Fig.~\ref{l1157}).
The abundances of methanol, ammonia, and a number of other molecules
are enhanced in B1 due to grain mantle evaporation~\cite{bach97}.
Only broad quasithermal lines were detected towards this region
both at 84 and 95~GHz~(Fig.~\ref{mas}).

In the region B2, about $30''$ offset from B1 (Fig.~\ref{l1157}),
the narrow feature at 7mm is very weak; moreover, its radial
velocity is the same as that of the narrow feature in B1.
Probably, there are no narrow lines, related to B2, and it is the
narrow feature related to B1 that fell into the antenna beam since
its width is as large as 82 arcsec\footnote{We cannot check this
suggestion using the ratio of observed intensities of the narrow
features in B1 and B2, since the accurate position of the source,
observed in B1 is unknown.}.

\section*{Discussion}
Typical widths of thermal lines towards regions of high-mass star
formation are about 3--5 km~s$^{-1}$ or higher. Therefore a
feature that was detected in a "maser" transition and is narrower
than 1~km~s$^{-1}$ can be considered a maser without any
additional justification. The situation in regions of
low-to-intermediate mass star formation is different. Widths of
thermal lines of quiescent gas in these regions are often 0.2--0.3
km~s$^{-1}$ or less. Therefore the fact that the detected line is
narrow does not necessarily mean that the line is a maser.
Additional interferometric observations for measuring the line
brightness temperature are necessary.

%

An essential argument in favor of maser origin of the detected
narrow features is the fact that these features are found in the
transitions $7_0-6_1A^+$ and $8_0-7_1A^+$ with the temperatures of
upper level excitation $E_u/k$ equal to 65~K and 84~K,
respectively. The kinetic temperature of quiescent gas in these
regions is about 10--20~K; therefore thermal excitation of these
transitions in quiescent gas is unlikely. These levels may be
thermally excited in the gas, heated and accelerated by shocks;
however, the width of thermal lines, arising in such gas, vary
within the range from several km~s$^{-1}$ to several dozen
km~s$^{-1}$.

In order to quantitatively verify this argument we modeled the
narrow features, found in NGC~1333IRAS4A, HH~25MMS, and L1157B1.
For this purpose, we computed a grid of LVG models spanning the
ranges 10--200~K in temperature, $0.32\times 10^4 - 1.0\times
10^8$~cm$^{-3}$ in density, and $0.1\times 10^{-4} - 0.56\times
10^{-1}$~cm$^{-3}$/(km~s$^{-1}$/pc) in methanol density
divided by the velocity gradient. We utilized the collisional
decay rates for methanol, obtained by
Pottage~et~al.~\cite{pottage} as a result of quantum mechanical
calculations. The agreement between the models and observational
data was estimated from the $\chi2$ criterion. The values of
$\chi2$ were calculated from the formula:
\begin{equation}
\chi2 = \sum_I\Biggl( \frac{R^I_{obs}-R^I_{mod}}{\sigma_{R^I_{obs}}}\Biggr)^2
\end{equation}
where $R^I_{obs}$ and $R^I_{mod}$ are the observed and model flux density
ratios for the narrow features in the $7_0-6_1A^+$, $8_0-7_1A^+$,
and $5_{-1}-4_0E$ lines, and $\sigma_{R^I_{obs}}$ are the errors of
the observed ratios. For HH~25MMS, we found that a number of models with
$T_{kin} \ge 95 K$ and \break
$5.6\times 10^4$~cm$^{-3} \le n_{H_2} \le 1.8\times 10^5$~cm$^{-3}$
satisfactorily reproduce the observed ratios. In all suitable models
the $5_{-1}-4_0E$, $8_0-7_1A^+$, and $7_0-6_1A^+$ lines proved to be inverted
with the absolute values of optical thickness of the $5_{-1}-4_0E$ and
$8_0-7_1A^+$ lines of the order of or higher than 3, and those of
the $7_0-6_1A^+$ line of the order of or higher than 5. The brightness
temperature of the $5_{-1}-4_0E$ and $8_0-7_1A^+$ lines in the suitable
models vary from several hundred to several thousand Kelvins, and that
of the $7_0-6_1A^+$ line is higher than 1000 K.

In the case of NGC~1333IRAS4A, agreement was obtained for a number
of models with $T_{kin} \ge 45$ K and $5.6\times 10^4$~cm$^{-3}\le n_{H_2}
\le 3.2\times 10^5$ cm$^{-3}$. The $5_{-1}-4_0E$, $8_0-7_1A^+$,
and $7_0-6_1A^+$ lines again proved to be inverted in all suitable
models. The absolute value of the optical thickness of these lines
varies within the range 3--5, and the brightness temperature
varies from several hundred to several thousand Kelvins.

We failed to find any suitable model for L1157B1, since the
intensities at 84 and 95~GHz that correspond to the intensity of
the narrow feature at 44~GHz are above the upper limits presented
in Table~\ref{gauss} in all computed models. There are several
explanations for this inconsistency within the frames of maser
hypothesis. For example, Sobolev~et~al.~\cite{sam25} suggested
that compact masers arise in turbulent media from the fact that in
a turbulent velocity field the coherence lengths along some
directions are larger than the mean coherence length. If the
masers are unsaturated, the intensity of maser emission is
proportional to $\exp(-\tau)$, where $\tau$ is the optical
thickness along the direction. In the first approximation one can
consider that the LVG modeling yields the opacity that is averaged
over all directions and the "mean" line brightness that
corresponds to this opacity. The larger the mean line opacity the
higher the ratio of the maximum brightness to the mean brightness.
Consider an example, when the coherence length along some
direction is twice its mean value. If the mean opacity equals
$-5$, which corresponds to the typical opacity of the $7_0-6_1A^+$
line, obtained as a result of the modeling of HH~25MMS, then the
ratio of the peak opacity to the mean opacity equals 2, and the
peak opacity equals $-10$. The ratio of the peak brightness
temperature (i.e. the brightness temperature of the maser spot) to
the mean brightness temperature (i.e. to the model brightness
temperature) equals $\exp(10)/\exp(5) \approx 150$. When the mean
opacity equals $-3$, which corresponds to the typical opacities of
the  $8_0-7_1A^+$ and $5_{-1}-4_0E$ lines, the ratio of the
brightness temperature of the maser spot to the model brightness
temperature equals  $\exp(6)/\exp(3) \approx 20$. Thus, in the
current example the ratio of the brightness temperatures of maser
spots in the $8_0-7_1A^+$ and $5_{-1}-4_0E$ lines to that of the
$7_0-6_1A^+$ line is less than the model ratio of the intensities
of these lines by a factor 7.5. This difference is sufficiently
large to explain the nondetection of maser features in the
$8_0-7_1A^+$ and $5_{-1}-4_0E$ lines in L1157B1. The increase of
the scattering of coherence lengths may lead to much greater
difference between the observed and model ratios of line intensities.

\begin{figure}[t]
\hspace{3cm}
\includegraphics{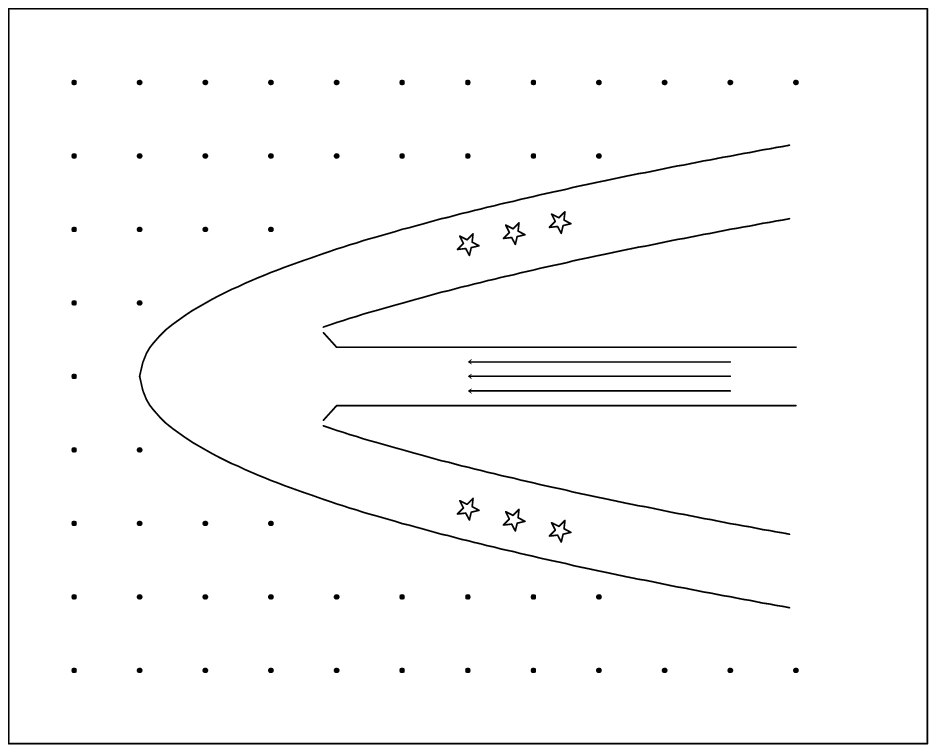}
\caption{A section of a bipolar outflow wing in the plane
perpendicular to the direction towards the observer. The stars
denote the possible positions of methanol masers. The
high-velocity jet (denoted by three arrows) and the shell that
appear behind the shock front are located between the solid lines.
The points denote the quiescent gas.} \label{shell}
\end{figure}

It is interesting that towards the region B2, located $30''$ from
B1, Kalenskii~et~al.~\cite{kalen01} observed in 2000 narrow
features at 95, and probably at 84~GHz, whereas in 2004 we found
at 7mm towards this region only a very weak narrow feature;
moreover, this feature may be related to B1 rather than to B2 (see
above). As far as the authors know, no Class I maser sources with
the emission at 44~GHz stronger than that at 84 and 95~GHz have
been found so far. Probably, Kalenskii~et~al. observed a variable
maser, whose intensity had essentially decreased by December 2004.
Variability of at least two Class I maser sources, W3(OH) and
G11.94-0.62, have been noted by Kurtz~et~al.~\cite{kurtz}. Another
possible explanation lies in the assumption that the maser at
44~GHz is linearly polarized and, during the observations in 2004,
its polarization plane was orthogonal to that of the receiver.
Linear polarization of the $7_0-6_1A^+$ masers has not been
measured so far; however, linear polarization has been detected by
Wiesemeyer~et~al.~\cite{polariz} in the $5_{-1}-4_0E$,
$8_0-7_1A^+$, and $6_{-1}-5_0E$ lines, the degree of the polarization in the
$6_{-1}-5_0E$ line being as large as 33\%--39\%. Probably,
the degree of the polarization in the stronger $7_0-6_1A^+$ line is
the same or even higher and may strongly affect the measured flux density.

Interestingly, the radial velocities of maser features virtually
coincide with the radial velocities of the peaks of thermal lines
$2_K-1_K$, which, in turn, are close to the radial velocities of
the quiescent gas. Gueth~et~al.~\cite{gueth96,gueth98} performed
interferometric observations of L1157 in CO and SiO lines. They
revealed hollow shells and related these shells to shock waves.
One can reasonably suggest that enhancement of methanol abundance
also occur in these shells. The brightest masers must arise in the
directions with the maximum opacity, i.e. along the walls, and
therefore must be observed towards the positions indicated in
Figure~\ref{shell}. Since the gas parameters apparently vary with
position, the maser region must occupy only a small part of the
shell. Since the outflow axis is virtually perpendicular to the
direction towards the observer, the radial velocity of maser
emission must be close to the radial velocity of the molecular
cloud. This model, as well as the aforementioned turbulent model,
suggests a preferential increase of brightness of the most intense
lines with respect to the brightness ratios for different lines,
obtained for a uniform source of the same temperature and density.
In order to test this model one should perform interferometric
observations at 44~GHz, first, of L1157, and to compare the
results with already available interferometric maps in other
molecular lines.

Thus, further interferometric observations are necessary both in
order to test whether the detected narrow lines are actually
masers and to choose between different maser models. The detection
of Class I methanol masers in relatively nearby regions of
low-mass star formation may have a strong impact on the
exploration of these masers. The regions of low-mass star
formation are in the whole much closer to the Sun and much better
studied than massive star-forming regions. Therefore it is much
easier to identify the observed masers with other objects (e.g.
with the wings of bipolar outflows) than when observing regions of
high-mass star formation.

\section*{Conclusion}
As a result of a short survey of bipolar outflows driven by
low-mass young stars in the methanol lines $7_0-6_1A^+$,
$8_0-7_1A^+$, and $5_{-1}-4_0E$ we detected narrow features
towards NGC~1333IRAS4A, HH~25MMS, and L1157~B1. Flux densities of
these features are no higher than 11~Jy, which is much lower than
the flux densities of strong maser lines in regions of high-mass
star formation. Nevertheless, most likely the narrow features are
masers. In order to confirm this suggestion further
interferometric observations are required; in addition, these
observations may help to make a choice between different maser
models.

{\em Acknowledgements} The authors are grateful to the staff of
Onsala Space Observatory for help during the observations. The
work was performed under partial financial support from the
Russian Foundation for Basic Research (grants Nos.~01-02-16902 and
04-02-17057) and the RAS Scientific Research Program "Extended Sources in
the Universe". Onsala Space Observatory is the Swedish National
Facility for Radio Astronomy and is operated by Chalmers
University of Technology, G\"{o}teborg, Sweden, with financial
support from the Swedish Research Council and the Swedish Board
for Technical Development.

\end{document}